\documentclass[10pt,a4paper]{article}
\usepackage{jheppub_kim}
\usepackage{pdflscape}
\usepackage{amsmath}
\usepackage{amssymb}
\usepackage{dcolumn}
\usepackage{bm}
\usepackage{color}
\usepackage{amsmath}
\usepackage{breqn}
\usepackage{graphicx}
\usepackage{epsfig}
\usepackage{caption,subcaption}
\usepackage{amsfonts}
\usepackage{graphicx}
\usepackage{parskip}
\usepackage{dcolumn}
\UseRawInputEncoding

\begin{document}

\title{Godel Universe in $f(Q,T)$ gravity: Exploring causality violation and closed time-like curves}

\author[a]{Tuhina Ghorui,}
\author[b]{Prabir Rudra,}
\author[c]{Farook Rahaman}

\affiliation[a] {Department of Mathematics, Jadavpur University, Kolkata-700 032, India.}

\affiliation[b] {Department of Mathematics, Asutosh College,
Kolkata-700 026, India.}

\affiliation[c] {Department of Mathematics, Jadavpur University, Kolkata-700 032, India.}

\emailAdd{tuhinaghorui.math@gmail.com}
\emailAdd{prudra.math@gmail.com}
\emailAdd{farookrahaman@gmail.com}

\abstract{In this work, the classical Godel solution from general relativity is extended into the framework of modified gravity theories based on non-metricity $Q$ and the trace of the energy-momentum tensor $T$ in the context of $f(Q,T)$ gravity. The main feature of the Godel solution is the existence of closed time-like curves, which allow for causality violation and time travel. Since general relativity and its extensions do not demand spacetime to be globally causal, there is good motivation to explore such solutions. We have found classes of solutions with different matter content, like perfect fluid, cosmological constant, massless scalar field, etc. It is observed that, for suitable initial conditions, there is always a possibility of obtaining feasible solutions that violate causality in our setup. The presence of non-metricity in such solutions produces crucial deviations that are noteworthy.}

\maketitle

\section{Introduction}
The perturbative investigations of quantum gravity that have demonstrated that general relativity (GR) is not renormalizable \cite{GA, ghor, GA2} are the primary source of motivation for changes in the framework of GR. Moreover, various astrophysical observations have demonstrated the universe's accelerating expansion \cite{Ae}. The most widely used changes are based on the addition of higher derivative terms, which are believed to enhance the fundamental renormalization features \cite{KP} of the theories. At the same time, changes of gravity that defy Lorentz and/or CPT garner a great deal of scholarly attention \cite{VP}. 

The late-time cosmic acceleration is being described by modified gravity theories, which have lately surfaced as an alternative to orthodox GR-based cosmology and are gaining popularity. Changing the conventional theory of general relativity to depict dark energy as a geometrical property of the universe is an interesting way to comprehend its nature. The alteration of the Einstein-Hilbert action is the foundation of these modified gravity theories \cite{mg1, mg2}. There are several fascinating modified theories of gravity, such as $f(R)$ gravity \cite{nab, nab1} and $f(R,T)$ gravity \cite{fg, fg1, fg2}. $f(T)$ Gravity \cite{ct, ct1}, sometimes referred to as teleparallel equivalent general relativity (TEGR), is another well-known modified theory of gravity, where the action substitutes the Ricci scalar $R$ with an arbitrary function of the torsion scalar $T$. Symmetric teleparallel gravity (STG) provides an alternative perspective to general relativity's conventional representations of curvature and torsion. This formalism creates $f(Q)$ gravity \cite{fq, fqq} by defining a geometric variable by a non-metricity scalar $Q$. To extend the $f(Q)$ theory, non-minimal coupling between the trace of the energy-momentum tensor and the gravitational interactions produced by the term $Q$ must be taken into account. As a result, an action is produced with an arbitrary function $f(Q,T)$ \cite{fqttt, fqttt1, fqttt2}. In $f(Q,T)$ gravity, the gravitational action is dependent on both the trace of the energy-momentum tensor $T$ and the non-metricity scalar $Q$. By adding matter-geometry coupling through the trace $T$, this theory expands on symmetric teleparallel gravity and may help explain dark energy, late-time cosmic acceleration, and other cosmological phenomena. This theory actually leads to modified gravitational dynamics in the absence of exotic matter. By effectively breaking the conservation of the energy–momentum tensor, this coupling offers novel explanations for dark energy behavior, cosmic acceleration, and large-scale deviations from General Relativity. Thus, the $f(Q,T)$ paradigm provides a rich field for studying early and late time cosmology, anisotropic universes, and the potential resolution of cosmic tensions by combining geometric extensions of gravity with phenomenological matter effects. Because of its versatility, it is a strong contender to investigate quantum-gravitational corrections and test the equivalence principle's bounds in non-Riemannian spacetime geometries.

Einstein’s equations describe how matter and energy curve spacetime, but they do not demand that spacetime must always be globally causal. So there is always motivation to explore such solutions that violate the global concept of causality. In general relativity, a class of solutions to Einstein's field equations known as Godel-type spacetimes has rotating and non-trivial causal structures, such as the potential for closed timelike curves (CTCs). In \cite{KG} Kurt Godel first proposed them in 1949 as a possible solution of general relativity. A closed timelike curve is a path in spacetime that is timelike at every point (meaning, a massive particle could in principle travel along it), and yet closes onto itself — returning to the same spacetime event. This means an observer following that trajectory would return to their own past — potentially meeting their younger self. Therefore, a CTC is a worldline that allows time travel — not through external machinery, but due to the curvature of spacetime itself. Time and space directions mix beyond a certain radius as a result of the global rotation of this spacetime, which "drags" light cones. Closed timelike loops occur through all points, and certain timelike pathways curve back on themselves. The universe lacks a universal concept of simultaneity since there is no global time function. As the first cosmic solution containing rotating matter, the Godel metric \cite{KG} is one of the most significant solutions in general relativity. This solution is cylindrically symmetric, stationary, and spatially homogeneous. Its highly nontrivial property is the breaking of causality, which suggests the possibility of closed timelike curves (CTCs) in Godel space, in contrast to Hawking's \cite{sw} conjecture that CTCs would not exist. Additionally, this metric was expanded to cylindrical coordinates in \cite{MJP, MJP1, MJP2}, where the causality issue was further investigated. As a result, it was discovered that three distinct classes of solutions could be distinguished. The following potential outcomes define these solutions: (i) There is just one noncausal region, (ii) no CTCs, and (iii) an endless series of alternating causal and noncausal regions. The quantities known as supermomentum and superenergy, which can be utilised as a criterion for the probability of existence of the CTCs, were introduced in the study presented in \cite{MDJ}. The string context is used in \cite{jbm, jbm1} to discuss the CTC solutions in the Godel spacetime (see also \cite{ofn} for a survey of various CTC issues). The possibility of nontrivially embedded black holes \cite{XSP} in the Godel spacetime is another factor that makes the Godel solution intriguing. In \cite{nn1} Godel solutions were studied in $f(Q)$ theory. Godel-type universes in the energy-momentum-squared gravity were studied in \cite{nn2}. Godel universe solutions in $f(R,T)$ gravity were studied in \cite{nn3}. Godel-type solutions in $f(R, T, R_{\mu\nu}T^{\mu\nu})$ gravity were explored in \cite{nn4}.

In this work, we would like to explore closed time-like curves in the background of $f(Q,T)$ gravity. Existence and implications of CTCs in modified gravity theories have been a fascinating topic of research for quite some time. It is expected that the effect of non-metricity and the coupled matter component will have a significant effect on the CTCs. A new method for investigating alternate cosmologies, such as $f(Q,T)$ gravity, can be investigated in rotating universes like the one offered by Godel's solution. It may be possible to test the consistency of $f(Q,T)$ gravity in the light of causality using the Godel metric. Since GR and its extensions do not demand spacetime to be globally causal, there is good motivation to explore such solutions. The paper is organized as follows: In Section 2, we discuss the details of $f(Q,T)$ gravity. In section 3, we study the intricacies of Godel-type spacetime, and in section 4, we extend the discussion to $f(Q,T)$ gravity. Section 5 is dedicated to the study of Godel-type solutions for perfect fluid and violation of causality. In section 6, we study the causal solutions with a massless scalar field. Finally, the paper ends with a discussion and conclusion in section 7.

\section{Overview of $f(Q,T)$ gravity}
 We start with the action of $f(Q,T)$ gravity  
\begin{equation}\label{t1}
S=\int \left[\frac{1}{2} f\left(Q,T\right) + \mathcal L_{m}\right] \sqrt{-g} d^{4}x
\end{equation}
where $g \equiv det (g_{\mu\nu})$ is the determinant of the metric tensor, Lagrangian of matter is $\mathcal L_{m}$, $T=g^{\mu\nu}T_{\mu\nu}$ is the trace of the stress-energy tensor, where $T_{\mu\nu}$, is defined as
\begin{equation}
T_{\mu\nu}=-\frac{2}{\sqrt{-g}}\frac{\delta\sqrt{-g}\mathcal{L}_m}{\delta g_{\mu\nu}}
\end{equation}
which can be further expressed as \cite{RIE}
\begin{equation}     
T_{\mu\nu}=g_{\mu\nu} -2 \frac{\delta \mathcal{L}_{m}}{\delta g^{\mu\nu}}
\end{equation}
The non-metricity tensor is given by
\begin{equation}
Q=Q_{\lambda\mu\nu} P^{\lambda\mu\nu}=-\frac{1}{2}Q_{\lambda\mu\nu}L^{\lambda\mu\nu}+\frac{1}{4}Q_{\lambda}Q^{\lambda}-\frac{1}{2}Q_{\lambda}\overset{\sim}{Q^{\lambda}}
\end{equation}
where the superpotential, is defined as
\begin{equation}
P^{\lambda}_{\mu\nu}=\frac{1}{4}\left(-2L^{\lambda}_{\mu\nu}+Q^{\lambda}\right)
\end{equation}
and the deformation tensor is defined by
\begin{equation}
L^{\alpha}_{\beta\gamma} \equiv -\frac{1}{2} g^{\alpha\lambda}\left[\nabla_{\nu} g_{\beta\lambda}+\nabla_{\beta} g_{\lambda\nu}-\nabla_{\lambda} g_{\beta\nu}\right]
\end{equation}

Taking the variation of the action in equation (\ref{t1}) we get
\begin{equation}
\delta S = \int \frac{1}{2}
\left[\delta f(Q,T )\sqrt{-g}\right] +\delta\left[\mathcal L_{m} \sqrt{-g}\right] d^{4}x
\end{equation}
which, on further simplification, gives
\begin{eqnarray*}
\delta S=\int \left[\frac{1}{2}\left(f_{Q}(Q,T) \delta Q+f_{T}(Q,T)\frac{\delta T}{\delta g^{\mu\nu}}\delta g^{\mu\nu}-\frac{1}{2}g_{\mu\nu}f(Q,T) \delta g^{\mu\nu}\right)\sqrt{-g}\right.
\end{eqnarray*}

\begin{equation}\label{vvv}
\left.+ \left(\sqrt{-g}\frac{\delta \mathcal L_{m}}{\delta g^{\mu\nu}}-\frac{\mathcal L_{m}}{2\sqrt{-g}}\frac{\delta g}{\delta g^{\mu\nu}}\right)\delta g^{\mu\nu} \right]d^{4}x
\end{equation}
The variation of the non-metricity scalar is given by
\begin{equation}
\delta Q= \left(P_{\mu\alpha\beta}Q_{\nu}^{\alpha\beta}-2Q^{\alpha\beta}_{\mu}P_{\alpha\beta\nu}+2P_{\alpha\mu\nu}\nabla{^\alpha}\right)\delta g^{\mu\nu}
\end{equation}

The energy-momentum tensor's variation with respect to the metric tensor is as follows
\begin{equation}\label{tgm}
\frac{\delta T}{\delta g^{\mu\nu}}=\frac{\delta g^{\alpha\beta}}{\delta g^{\mu\nu}} T_{\alpha\beta} + g^{\alpha\beta}\frac{\delta \Gamma_{\alpha\beta}}{\delta g^{\mu\nu}}
\end{equation}
Further we define
\begin{equation}\label{tht}
\Theta_{\mu\nu} \equiv g^{\alpha\beta} \frac{\delta \Gamma_{\alpha\beta}}{\delta g^{\mu\nu}}
\end{equation}

Continuing from eqn.(\ref{vvv}) we get
\begin{eqnarray*}
\delta S = \int \left[\frac{1}{2}\left\{f_{Q}(Q,T)\left(P_{\mu\alpha\beta}Q_{\nu}^{\alpha\beta}-2Q^{\alpha\beta}_{\mu}P_{\alpha\beta\nu}+2P_{\alpha\mu\nu}\nabla^{\alpha}\right)+f_{T}(Q,T)(T_{\mu\nu}+\Theta_{\mu\nu})\right.\right.
\end{eqnarray*}
\begin{equation}
\left. \left.-\frac{f(Q,T)}{2}g_{\mu\nu}\right\}-\frac{T_{\mu\nu}}{2}\right] \sqrt{-g}\delta g^{\mu\nu} d^{4}x
\end{equation}

The field equations are obtained as
\begin{equation}
\frac{2}{\sqrt{-g}} \nabla_{\alpha}\left(f_{Q}\sqrt{-g}P^{\alpha}_{\mu\nu}\right)-\frac{1}{2}f(Q,T)g_{\mu\nu}+f_{T}\left(T_{\mu\nu}+\Theta_{\mu\nu}\right)+f_{Q}(Q,T)\left(P_{\mu\alpha\beta}Q_{\nu}^{\alpha\beta}-2Q^{\alpha\beta}_{\mu}P_{\alpha\beta\nu}\right)= T_{\mu\nu}
\end{equation}
where the partial derivative of $f$ with respect to $Q$ and $T$ are indicated by $f_{Q}$ and $f_{T}$ respectively.

From the above equation, we derive,
\begin{equation}
\frac{2}{\sqrt{-g}} \nabla_{\alpha}(f_{Q}\sqrt{-g}P^{\alpha}_{\mu\nu})+f_{Q}(Q,T)\left(P_{\mu\alpha\beta}Q_{\nu}^{\alpha\beta}-2Q^{\alpha\beta}_{\mu}P_{\alpha\beta\nu}\right)=f_{Q} \overset{\circ}{G_{\mu\nu}}+\frac{1}{2}f_{Q}Q g_{\mu\nu}+f_{QQ}\partial_{\alpha}QP^{\alpha}_{\mu\nu}
\end{equation}

Finally, the field equation of $f(Q, T)$ gravity can be rewritten as
\begin{equation}\label{tr}
f_{Q} \overset{\circ}{G_{\mu\nu}}+\frac{1}{2}(Qf_{Q}-f)g_{\mu\nu}+f_{QQ}\partial_{\alpha}Q P^{\alpha}_{\mu\nu}= T_{\mu\nu} -f_{T}\left(T_{\mu\nu}+\Theta_{\mu\nu}\right)
\end{equation}
where $\overset{\circ}{G_{\mu\nu}}=R_{\mu\nu}-\frac{1}{2}Rg_{\mu\nu}$ is the Einstein tensor connected to the Levi-Civita connection.

From equation (\ref{tgm}) we get,
\begin{equation}\label{t3}
\frac{\delta T_{\alpha\beta}}{\delta g^{\mu\nu}}=L_{m}\frac{\delta g_{\alpha\beta}}{\delta g^{\mu\nu}}+\frac{1}{2}g_{\alpha\beta}g_{\mu\nu}L_{m}-\frac{1}{2}g_{\alpha\beta}T_{\mu\nu}-2\frac{\delta^{2}L_{m}}{\delta g^{\alpha\beta}\delta g^{\mu\nu}}
\end{equation}
From equation (\ref{tht}) we have,
\begin{equation}
\Theta_{\mu\nu}=-2T_{\mu\nu}+g_{\mu\nu}L_{m}-2g^{\alpha\beta}\frac{\partial^{2}L_{m}}{\partial g^{\alpha\beta} \partial g^{\mu\nu}}
\end{equation}
Assuming that the matter Lagrangian is given  $\mathcal{L}_{m}=-p$, we get
\begin{equation}
\Theta_{\mu\nu}=-2T_{\mu\nu}-pg_{\mu\nu} 
\end{equation}

In the subsequent sections we will use the above expressions to derive Godel-type solutions in $f(Q,T)$ gravity and discuss their properties and implications in cosmology.

\section{Godel-type spacetime}
Kurt Godel's 1949 cosmological solution to Einstein's field equations remains a significant source of inspiration for further research into more complex solutions. We begin our study by introducing the Godel line element \cite{godd1}
\begin{equation}
ds^{2}=a^{2}\left[dx_{0}+H(x_{1})dx_{2}\right]^{2}-D^{2}(x_{1})dx_{2}^{2}-dx_{1}^{2}-dx_{3}^{2}
\end{equation}
with $H=e^{mx_{1}}$ and $D=\frac{e^{mx_{1}}}{\sqrt{2}}$. With these values, the metric can be rewritten as,
\begin{equation}\label{t2}
ds^{2}=a^{2}\left(dx_{0}^{2}-dx_{1}^{2}+\frac{e^{2mx_{1}}}{2}dx_{2}^{2}-dx_{3}^{2}+2e^{mx_{1}}dx_{0}dx_{2}\right)
\end{equation}
where $a$ is a positive number compatible with the model. This is a homogeneous and rotating spacetime. The necessary and sufficient conditions for this metric to be spacetime homogeneous are
\begin{equation}\label{hd}
\frac{H'}{D}=\sqrt{2}m=2\omega,  ~~~~~ \frac{D''}{D}=m^2
\end{equation}
where dash denotes derivatives. The parameter $\omega$ is the rate of rigid rotation of matter (vorticity) and $m$ is a constant, such that $\omega$ is real and non-zero while $-\infty<m^{2}<\infty$. Consequently, there are three types of possible Godel-type metrics (i) hyperbolic class ($m^{2}>0$), (ii) trigonometric class ($m^{2}<0$),  (iii) linear class ($m^{2}=0$). In the Godel spacetime, due to the presence of the off-diagonal term $2e^{mx_{1}} dx_{0}dx_{2}$, the coordinate $x_{2}$ (which is periodic) becomes mixed with the coordinate $x_{0}$, which represents time. This implies that an observer can return to the same spacetime position with the same value of $x_0$ or even earlier by travelling far enough in the $x_2$-direction (around an axis). Mathematically, for fixed $x_{1}>x_{1c}$, the line $x_{0}=constant, x_{1}=constant, x_{3}=constant$ is timelike. Plugging $dx_{0}, dx_{1}, dx_{3}=0$ in the metric we get $ds^{2}=\frac{1}{2}a^{2}e^{2mx_{1}}dx_{2}^{2}$. At small values of $x_{1}$, this term is spacelike, but beyond a certain critical value $x_{1c}$, the direction along $x_{2}$ becomes timelike. Thus, the motion around the axis becomes a timelike loop-a closed timelike curve. From a physical point of view, Gödel's universe is a rotating cosmological model, where the spacetime itself rotates rather than matter in space. This rotation twists the light cones until they topple over by "dragging" local inertial frames around (a powerful gravitomagnetic effect).

The components of the metric tensor $g_{\mu\nu}$ are
\begin{equation}
g_{\mu\nu}=a^{2}
\begin{pmatrix}
  1 & 0 & e^{mx_{1} } & 0\\
  0 & -1 & 0 & 0\\
  e^{mx_{1}} & 0 & \frac{1}{2}e^{2mx_{1}} & 0\\
  0 & 0 & 0 & -1
    
\end{pmatrix}   
\end{equation}
The inverse metric tensor is given by
\begin{equation}
g^{\mu\nu}=\frac{1}{a^{2}}
\begin{pmatrix}
  -1 & 0 & 2e^{-mx_{1}} & 0\\
  0 & -1 & 0 & 0\\
  2e^{-mx_{1}} & 0 & -2e^{-2mx_{1}} & 0\\
  0 & 0 & 0 & -1
    
\end{pmatrix}   
\end{equation}

The non-zero Christoffel symbols for this metric are as follows:

$\Gamma_{01}^{0}=m,~ \Gamma_{12}^{0}=\Gamma_{02}^{1}=\frac{me^{mx_{1}}}{2},~ \Gamma_{22}^{1}=\frac{me^{2mx_{1}}}{2},~ \Gamma_{01}^{2}=-me^{-mx_{1}}$

The following expression provides the Ricci curvature tensor
\begin{equation}
R_{ik}=\frac{\partial}{\partial x_{\sigma}}\Gamma^{\sigma}_{ik}-\frac{1}{2}\frac{\partial^{2}log (g)}{\partial x_{i}\partial x_{k}}+\frac{1}{2}\Gamma^{\sigma}_{ik}\frac{\partial log (g)}{\partial x_{\sigma}}-\Gamma^{\rho}_{\sigma i}\Gamma^{\sigma}_{\rho k}
\end{equation}
Here $\frac{\partial}{\partial x_{\mu}}=0$ vanishes, except for $\mu = 1$. Moreover $g=det (g_{\mu\nu})=\frac{a^{8}}{2} e^{2x_{1}}$, therefore $log(g)$ is linear in $x_{1}$.

The nonzero components of the Ricci tensor are,
\begin{equation}
R_{00}=m^{2}, R_{22}=m^{2}e^{2mx_{1}}, R_{02}=R_{20}=m^{2}e^{mx_{1}}
\end{equation}

The Ricci scalar is given by, 
\begin{equation}
R=\frac{m^{2}}{a^{2}}
\end{equation}
Assuming that four-velocity vector $u^{\mu}$ as $u^{\mu} = \left(\frac{1}{a}, 0, 0, 0\right)$ and $u_{\mu}=(a, 0, ae^{mx_{1}}, 0),$ 
The Einstein tensor is given by,
\begin{equation}
\overset{\circ}{G_{\mu\nu}}=R_{\mu\nu}-\frac{1}{2}Rg_{\mu\nu}
=8\pi \rho u_{\mu}u_{\nu}+\Lambda g_{\mu\nu}
\end{equation}
Using the transformations, $\theta^{0}=a\left(dx_{0}+e^{mx_{1}}dx_{2}\right)$, $\theta^{1}=a dx_{1}$, $\theta^{2}=a\left(\frac{e^{mx_{1}}}{\sqrt{2}}dx_{2}\right)$, $\theta^{3}=adx_{3}$ the metric (\ref{t2})can be rewritten in the form 
\begin{equation}\label{th}
ds^{2}=\eta_{\alpha\beta}\theta^{\alpha}\theta^{\beta} =(\theta^{0})^{2}-(\theta^{1})^{2}-(\theta^{2})^{2}-(\theta^{3})^{2}
\end{equation} 
where, $\eta_{\alpha\beta}= diag(1,-1,-1,-1)$.
  
The non-vanishing components of the tetrad $e^{(\alpha)}_{\mu}$ are
\begin{equation}\label{th1}
e^{(0)}_{0}=a;~ e^{(1)}_{1}=a;~ e^{(0)}_{2}=ae^{mx_{1}};~ e^{(2)}_{2}=\frac{a}{\sqrt{2}}e^{mx_{1}};~ e^{(3)}_{3}=a.
\end{equation}

The non-zero component of the inverse tetrad
$e^{(\mu)}_{\beta}$, defined by
$e^{\alpha}_{\mu}e^{\mu}_{\beta}=\delta^{\alpha}_{\beta}$ are given as
\begin{equation}\label{th2}
e^{0}_{(0)}=e^{1}_{(1)}=e^{3}_{(3)}=\frac{1}{a};~~ e^{0}_{(2)}=-\frac{\sqrt{2}}{a};~~
e^{2}_{(2)}=\frac{\sqrt{2}}{a}e^{-mx_{1}}.
\end{equation}

\section{Godel-type metric in $f(Q,T)$ gravity}
Here we are primarily interested in investigating if Godel-like violations of causality are admitted in $f(Q,T)$ gravity.  Therefore, we need to find out if $f(Q,T)$ gravity admits any feasible Godel-type solutions. We use the coincident gauge for finding solutions to the field equations. The definition of the coincident gauge is
\begin{equation}
\Gamma^{\mu}_{\alpha\beta}=0
\end{equation}
which implies 
\begin{equation}
\nabla_{\alpha}=\partial_{\alpha}
\end{equation}
i.e., the covariant derivative is the same as the usual partial derivative. The nonmetricity components in this gauge are 
\begin{equation}
    Q_{\mu\alpha\beta}=\partial_{\mu}g_{\alpha\beta}
\end{equation}
The non-vanishing component of the nonmetricity tensor corresponding to the metric (\ref{t2}) are

\begin{equation}
      Q_{102}=Q_{120}=a^{2}m e^{mx_{1}};~ Q_{122}=a^{2}m e^{2mx_{1}}
\end{equation}

The trace of the nonmetricity tensor is defined by
\begin{equation}
Q_{\alpha}=g^{\mu\nu}Q_{\alpha\mu\nu}
\end{equation}
The non-vanishing components of $Q_{\alpha}$ are given by
\begin{equation}
Q_{1}=2m
\end{equation}
The components of $Q^{\alpha}$ are given by 
\begin{equation}
Q^{\alpha}=g^{\alpha\beta}Q_{\beta}
\end{equation}
From the above we get,
\begin{equation}
Q^{1}=\frac{2m}{a^{2}}
\end{equation}

The non-metricity scalar can be written as 
\begin{equation}
 Q=\frac{1}{4}Q_{\alpha}Q^{\alpha}=\frac{m^{2}}{a^{2}}=2\omega^{2}
\end{equation}
The field equations for the metric (\ref{th}) have the following form in the tetrad basis defined by (\ref{th1}) and (\ref{th2}) 

\begin{equation}
f_{Q} \overset{\circ}{G_{AB}}+\frac{1}{2}(Qf_{Q}-f)\eta_{AB}+f_{QQ}\partial_{\alpha}Q P^{C}_{AB}=T_{AB} -f_{T}\left(T_{AB}+\Theta_{AB}\right)
\end{equation}
Where $\overset{\circ}{G_{AB}}=e^{\mu}_{A}e^{\nu}_{\beta}\overset{\circ}{G_{\mu\nu}},~ U_{C}=e^{\lambda}_{C}\partial_{\lambda}Q,~ P^{C}_{AB}=e^{C}_{\lambda}e^{\mu}_{A}e^{\nu}_{\beta}P^{\lambda}_{\mu\nu},~ T_{AB}=e^{\mu}_{A}e^{\nu}_{\beta}T_{\mu\nu}$.

Since $Q$ is constant, $U_{C}=0$. The field equations are reduced to
\begin{equation} \label{t4}
f_{Q} \overset{\circ}{G}_{AB}+\frac{1}{2}(Qf_{Q}-f)\eta_{AB}=T_{AB} +f_{T}\left(T_{AB}+p\eta_{AB}\right)
\end{equation}
The Einstein tensor's nonzero components in the tetrad basis are
\begin{equation}
\overset{\circ}{G}_{(0)(0)} = \overset{\circ}{G}_{(1)(1)}  = \overset{\circ}{G}_{(3)(3)} = \frac{m^{2}}{2a^{2}};~ \overset{\circ}{G}_{(2)(2)}=\frac{3m^{2}}{4a^{2}}
\end{equation}
Now we have all the necessary tools to discuss the physical implications of the solutions of the Godel metric in $f(Q,T)$ gravity. Now we will consider some particular toy models and discuss the feasibility, compatibility, and implications of the Godel solution.

\subsection{Model-1}
In the first model, we consider a minimal form of interaction between the non-metricity and matter through an analytical form $f(Q,T)=f_{1}(Q)+f_{2}(T)$. The energy-momentum tensor is taken as,
\begin{equation}
T_{\mu\nu}=8\pi\rho u_{\mu} u_{\nu}+\Lambda g_{\mu\nu}
\end{equation}
where the four-velocity vector $u_{\mu}$ is given by $u_{\mu}=(a,0,ae^{mx_{1}},0)$ and $\Lambda$ is the cosmological constant. The non-zero components of the energy-momentum tensor are given by,
\begin{equation}
T_{00}=(8\pi\rho+\Lambda) a^{2},~~ T_{02}=(8\pi\rho+\Lambda) a^{2}e^{m x_{1}},~~ T_{11}=-\Lambda a^{2},~~ T_{22}=(8\pi\rho+\frac{\Lambda}{2}) a^{2}e^{2mx_{1}},~~ T_{33}=-\Lambda a^{2}.
\end{equation}
The trace of the energy-momentum tensor is given by
\begin{equation}\label{tt2}
T=8\pi \rho+4\Lambda
\end{equation}
Here, the field equations (\ref{tr}) become,
\begin{equation}
f_{1}'(Q)  \overset{\circ}{G_{\mu\nu}} = T_{\mu\nu}- f'_{2}(T)(T_{\mu\nu}+\Theta_{\mu\nu})+\frac{1}{2} \left[f_{1}(Q) -Q f_{1}'(Q)+f_{2}(T)\right] g_{\mu\nu}+f_{1}''(Q) \partial_{\alpha}Q P^{\alpha}_{\mu\nu}
\end{equation}
where the prime represents a derivative with respect to the argument. To simplify, we assume that the matter is a perfect fluid with zero pressure $(p = 0)$. We have $\Theta_{\mu\nu}=-2T_{\mu\nu}$, and the field equations become
\begin{equation}
f_{1}'(Q) \overset{\circ}{G_{\mu\nu}} = f'_{2}(T)T_{\mu\nu}+ T_{\mu\nu} +\frac{1}{2} \left[f_{1}(Q)-Qf'_{1}(Q) +f_{2}(T)\right] g_{\mu\nu}+f_{1}''(Q) \partial_{\alpha}Q P^{\alpha}_{\mu\nu}
\end{equation}
This equation can be reformulated as an effective Einstein's field equation as
\begin{equation}
\overset{\circ}{G_{\mu\nu}}=\frac{1}{ f_{1}'(Q)}\left[1+f'_{2}(T)\right]T_{\mu\nu}+\frac{1}{ f_{1}'(Q)}\left[\frac{1}{2}(f_{1}(Q)-Qf'_{1}(Q)+f_{2}(T)]g_{\mu\nu}-2\left(f_{QQ}\nabla_{\lambda}Q\right)\right]P^{\lambda}_{\mu\nu}
\end{equation}
Let us consider the particular forms $f_{1}(Q)=Q$ and $f_{2}(T)=2\lambda T$, where $\lambda$ is constant. For this choice, the field equations become
\begin{equation}
\overset{\circ}{G_{\mu\nu}}=\left(1+2\lambda\right)T_{\mu\nu}+\lambda T g_{\mu\nu} 
\end{equation}
The non-zero components of the field equations are as follows:\\
For $(00)$ and $(02)$ components we get
\begin{equation}
\frac{m^{2}}{2 a^{2}}=a^{2}\left[8 \pi \rho(1+3\lambda)+\Lambda(1+6\lambda)\right]  
\end{equation}
For $(11)$ and $(33)$ components we get
\begin{equation}
\frac{m^{2}}{2a^{2}}=-a^{2}\left[8\pi\rho\lambda +\Lambda\left(1+6\lambda\right)\right]
\end{equation}
Similarly for the $(22)$ component we have,
\begin{equation}
\frac{3 m^{2}}{4 a^{2}}=a^{2}\left[8\pi\rho\left(1+\frac{5\lambda}{2}\right)+\frac{\Lambda}{2}\left(1+6\lambda\right)\right]
\end{equation}

The solution of these field equations is obtained as
\begin{equation}
\rho=\frac{m^{2}}{8\pi a^{4}(1+2\lambda)},~~~~~~~~
\Lambda=-\frac{(1+4\lambda)m^{2}}{2 a^{4}(1+2\lambda)(1+6\lambda)}
\end{equation}
When we put $\lambda=0$, we recover a solution comparable to the original solution of Godel \cite{KG}. The deviation occurs due to the contributions from the non-metricity. We can see that for $\lambda>0$ we get a positive value of the energy density $\rho$. The energy density of the universe is of the order of $10^{-27}$ $kg/m^{3}$. From the obtained solution, we see that by adjusting the parameters, it is possible to realize the observed value. Similarly, the observed value of the cosmological constant is of the order of $10^{-52}$ $m^{-2}$, which is also clearly achievable by fine-tuning the parameters. This shows that the solution obtained is feasible, subject to the fine-tuning of the parameter space. The existence of a feasible solution shows the possibility of a Godel universe in $f(Q,T)$ gravity.

\subsection{Model-II}
In the second model, we consider a non-minimal interaction between $Q$ and $T$ through the form $f(Q,T)=f_{1}(Q)+f_{2}(Q)f_{3}(T)$. For this set-up, the field equations (\ref{tr}) for a perfect fluid (dust) with $p=0$ are given by
\begin{eqnarray*}
\left(f_{1}'(Q)+f_{2}'(Q)f_{3}(T)\right) \overset{\circ}{G_{\mu\nu}}=\left[1+f_{2}(Q)f'_{3}(T)\right]T_{\mu\nu}
\end{eqnarray*}
\begin{equation}
+\left[\frac{1}{2}\left(f_{1}(Q)-Q\left(f'_{1}(Q)+f'_{2}(Q)\right)+f_{2}(Q)f_{3}(T)\right)g_{\mu\nu}-2\left((f_{1}''(Q)+f_{2}''(Q)f_{3}(T)\nabla_{\lambda}Q\right)\right]P^{\lambda}_{\mu\nu}
\end{equation}
Here we take $f_{1}(Q)=f_{2}(Q)=Q$ and $f_{3}(T)=\lambda T$.
Using the Godel metric and assuming a constant trace of the energy-momentum tensor (\ref{tt2}), we can simplify the field equations to
\begin{equation}
\overset{\circ}{G_{\mu\nu}}=\frac{1}{1+\lambda T}\left(8\pi+\lambda Q\right) T_{\mu\nu}
\end{equation}
where $Q=\frac{m^{2}}{a^{2}}$.
The nonzero components of the field equations are (00), (11), and (22), respectively, given by
\begin{equation}
\left(1+\lambda \left(8 \pi \rho+4 \Lambda\right)\right)\frac{m^{2}}{2a^{2}}=\left(8\pi+\lambda\frac{m^{2}}{a^{2}}\right)(8\pi\rho+\Lambda) a^{2}
\end{equation}
\begin{equation}
\left(1+\lambda (8 \pi \rho+4 \Lambda)\right)\frac{m^{2}}{2a^{2}}=-((8\pi+\lambda\frac{m^{2}}{a^{2}}))(\Lambda a^{2})
\end{equation}
\begin{equation}
\left(1+\lambda \left(8 \pi \rho+4 \Lambda\right)\right)\frac{3m^{2}}{4a^{2}}=\left(8\pi+\lambda\frac{m^{2}}{a^{2}}\right)\left(8\pi\rho+\frac{\Lambda}{2}\right)a^{2}
\end{equation}
The solution of the above system is obtained as
\begin{equation}
\rho=\frac{m^{2}}{8\pi \left(8a^{4}\pi+m^{2}\lambda+a^{2}m^{2}\lambda\right)},~~~~~~~\Lambda=-\frac{m^{2}}{2(8a^{4}\pi+m^{2}\lambda+a^{2}m^{2}\lambda)}
\end{equation}
When we put $\lambda=0$, we recover a solution comparable to the original solution of Godel \cite{KG}. The deviation occurs due to the contributions from the non-metricity. Just like the previous case, here also see that by adjusting the parameter space, we can realize the observed values of $\rho$ and $\Lambda$.

\section{Perfect fluid and Violation of Causality}
In this section, we will consider matter in the form of a perfect fluid and explore the Godel solutions in the background modified gravity. The energy-momentum tensor (EMT) of a perfect fluid is given by
\begin{equation}
T_{AB}=(\rho+p)u_{A}u_{B}-p\eta_{AB}
\end{equation}
where $u_{A}=(1,0,0,0)$.
The nonzero components of the EMT are 
\begin{equation}
T_{00}=\rho ,~~ T_{11}=T_{22}=T_{33}=p 
\end{equation}
For space-time homogeneous Godel-type metrics, the non-vanishing Lorentz frame components of the Einstein tensor $G_{AB}$ have the extremely basic form \cite{l1}
\begin{equation}
\overset{\circ}{G}_{(0)(0)} =3\omega^{2}-m^{2} ;~\overset{\circ}{G}_{(1)(1)}=\overset{\circ}{G}_{(3)(3)} = \omega^{2};~ \overset{\circ}{G}_{(2)(2)}=m^{2}-\omega^{2}
\end{equation}
The nontrivial elements of the field equations (\ref{t4}) for f(Q,T) gravity in this instance have the following structure:
\begin{equation}\label{t5}
(3\omega^{2}-m^{2})f_{Q}+\frac{1}{2}(Qf_{Q}-f)=(1+f_{T})\rho+pf_{T}
\end{equation}
\begin{equation}\label{t6}
\omega^{2}f_{Q}-\frac{1}{2}(Qf_{Q}-f)=p
\end{equation}
\begin{equation}\label{t7}
(m^{2}-\omega^{2}) f_{Q}-\frac{1}{2}\left(Qf_{Q}-f\right)=p
\end{equation}
The first equation represents the (00)-component, and the second equation represents the (11), (22), and the third equation (33) components.
from (\ref{t6}) and (\ref{t7}), we get
\begin{equation}
m^{2}=2\omega^{2}
\end{equation}

From (\ref{t5}) and (\ref{t6}) we get
\begin{equation}\label{ttt5}
\rho=\left[\frac{1-f_{T}}{1+f_{T}}\omega^{2}f_{Q}+\frac{1}{2}\left(Qf_{Q}-f\right)\right]
\end{equation}
and
\begin{equation}\label{ttt6}
p=\frac{1}{2} \left(\omega^{2}\left(f_{Q}-Qf_{Q}\right)+f\right)
\end{equation}
\begin{equation}
    \rho+p=\frac{2\omega^{2 }f_{Q}}{1+f_{T}}
\end{equation}
 Generally if $0<m^{2}<4\omega^{2}$, there is a radius beyond which causality is breached \cite{KG, MJP, MJR}. This critical radius is given by \cite{nn1}
\begin{equation}\label{bm}
x_{1c}=\frac{2}{\left|m \right|} sinh^{-1}\left(\frac{4\omega^{2}}{m^{2}}-1\right)^{-1}
\end{equation}
Therefore, considering eqn.(\ref{hd}) we get from the above expression,
\begin{equation}
x_{1c}=\frac{\sqrt{2}}{\left|\omega \right|}sinh^{-1}\left(1\right)=\frac{\sqrt{2}}{\left|\omega \right|}\ln{(1+\sqrt{2})}    
\end{equation}
There is a breach of causality when $x_{1}>x_{1c}$. This is because the circles $x_{1}=constant>x_{1c}$, $x_{0}=constant$, $x_{3}=constant$ are closed time-like curves \cite{KG, MJP, MJR}.

\subsection{Energy Conditions for perfect fluid matter}
In essence, energy conditions are assertions that matter cannot travel faster than light and that the total energy density is nonnegative throughout spacetime \cite{JMS, en1, en2}. To rule out unphysical solutions of the Einstein field equations, they should be obeyed by non-gravitational fields and normal matter. In the backdrop of $f(Q,T)$ gravity, let us investigate the degree to which the function $f(Q,T)$ is constrained by energy conditions on the ideal fluid that serves as the matter source of Godel's solution.
From (\ref{t5}) and (\ref{t6}) we get
\begin{equation}\label{ttt5}
\rho=\frac{1-f_{T}}{1+f_{T}}\omega^{2}f_{Q}+\frac{1}{2}\left(Qf_{Q}-f\right)
\end{equation}
and
\begin{equation}\label{ttt6}
p=\omega^{2}f_{Q}-\frac{1}{2}\left(Qf_{Q}-f\right)
\end{equation}
\begin{equation}
    \rho+p=\frac{2\omega^{2 }f_{Q}}{1+f_{T}}
\end{equation}
\begin{equation}
    \rho+3p=\left(\frac{2\left(2+f_{T}\right)}{1+f_{T}}\omega^{2}-Q\right)f_{Q}+f
\end{equation}

The weak energy condition is given by $\rho \geq 0$ and $\rho+p\geq0$. From equation (\ref{ttt5}) and (\ref{ttt6}) we have
\begin{equation}
\left[\frac{1-f_{T}}{1+f_{T}}\omega^{2}f_{Q}+\frac{1}{2}\left(Qf_{Q}-f\right)\right]\geq0
\end{equation}
and 
\begin{equation}
\frac{2\omega^{2}f_{Q}}{1+f_{T}}\geq 0
\end{equation} respectively.
For the strong energy condition $\rho+3p\geq0$ , we get
\begin{equation}
\left(\frac{2\left(2+f_{T}\right)}{1+f_{T}}\omega^{2}-Q\right)f_{Q}+f\geq 0
\end{equation}

According to the dominant energy condition,  $\rho-|p|\geq 0$
\begin{equation}
\rho-|p|=\frac{1-f_{T}}{1+f_{T}}\omega^{2}f_{Q}+\frac{1}{2}\left(Qf_{Q}-f\right)-|\omega^{2}f_{Q}-\frac{1}{2}\left(Qf_{Q}-f\right)|
\end{equation}
and if $p> 0$, then we get
\begin{equation}
\rho-|p|=\frac{2\left(1-f_{T}\right)}{1+f_{T}}\omega^{2}f_{Q}+\left(Qf_{Q}-f\right)
\end{equation}
 if $P< 0$, then this equation is reduced to 
\begin{equation}
\rho-|p|=\frac{2\omega^{2}f_{Q}}{1+f_{T}}
\end{equation}

This two inequality is similar to the one obtained for condition of the weak energy condition.
So, effectively, for the dominant energy condition, we get

Combining all the energy conditions, we see that the following inequalities must be satisfied simultaneously:
\begin{equation}\label{td}
\frac{2\left(1-f_{T}\right)}{1+f_{T}}\omega^{2}f_{Q}+Qf_{Q}-f \geq 0;~~ \frac{2\omega^{2}f_{Q}}{1+f_{T}}\geq 0 ;~~ \left(\frac{2\left(2+f_{T}\right)}{1+f_{T}}\omega^{2}-Q\right)f_{Q}+f\geq 0
\end{equation}
Now we consider two different toy models to illustrate the validity of the above energy conditions. 

\subsubsection{Model-1}
We consider the following functional form
\begin{equation}
f(Q,T)=\alpha Q^{n}+\beta T^{m_{1}}
\end{equation}
where $\alpha$, $\beta$, $m_{1}$ and $n$ are constants. Now, the inequalities in (\ref{td}) are reduced to
\begin{equation}
f_{1}(Q,T)=-T^{m_{1}}\beta+Q^{n}\alpha\left[-1+n\left(1-\frac{2\omega^{2}}{Q}+\frac{4T\omega^{2}}{Q(T+m_{1}T^{m_{1}}\beta)}\right)\right]\geq 0
\end{equation}
\begin{figure}[hbt!]
\begin{center}
\includegraphics[height=3in]{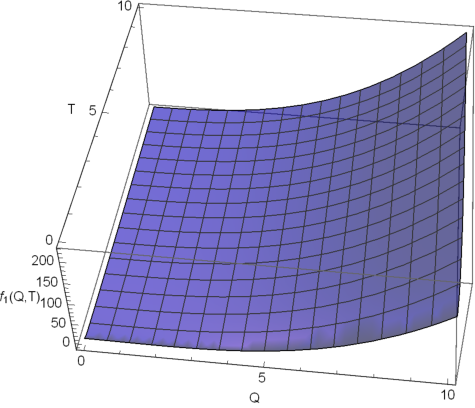}
\caption{The figure shows the plot of $f_{1}(Q,T)$ against $Q$ and $T$ for Model-1. The parameters are taken as $\alpha=0.1, \beta=7, n=3, m_{1}=0.2$ .}
\label{figscale1}
\end{center}
\end{figure}
\begin{equation}
f_{2}(Q,T)=\frac{2\omega^{2}\alpha n Q^{n-1}}{1+m_{1}\beta T^{m_{1}-1}}\geq 0
\end{equation}
\begin{figure}[hbt!]
\begin{center}
\includegraphics[height=3in]{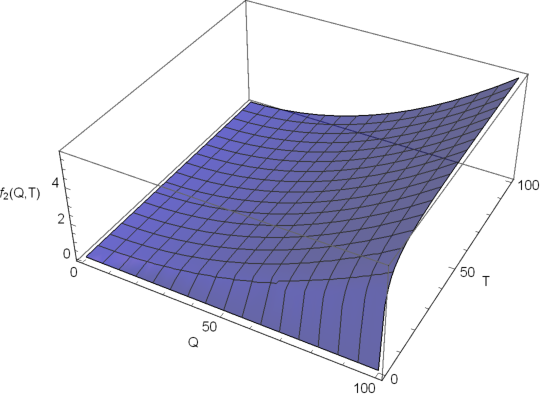}
\caption{The figure shows the plot of $f_{2}(Q,T)$ against $Q$ and $T$ for Model-1. Other parameters are given by $\alpha=0.1, \beta=7, n=3, m_{1}=0.2$ .}
\label{figscale1}
\end{center}
\end{figure}
\begin{equation}
f_{3}(Q,T)=T^{m_{1}}\beta+Q^{n}\alpha\left(1+n\left(-1+2\left(\frac{1}{Q}+\frac{1}{Q+m_{1}QT^{-1+m_{1}}\beta}\right)\omega^{2}\right)\right)\geq 0
\end{equation}
\begin{figure}[hbt!]
\begin{center}
\includegraphics[height=4in]{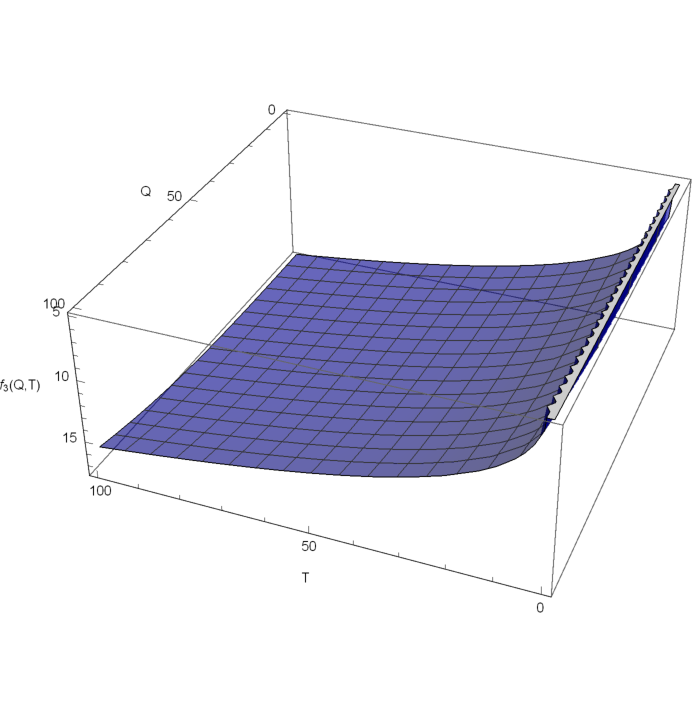}
\caption{The figure shows the plot of $f_{3}(Q,T)$ against $Q$ and $T$ for Model-1. Other parameters are given by $\alpha=0.1, \beta=7, n=3, m_{1}=0.2$ .}
\label{figscale1}
\end{center}
\end{figure}

Now we expect that there exists a parameter space for which the above inequalities are satisfied, thus validating the energy conditions. To illustrate this, we generate plots for the functions on the left-hand side of the inequalities, i.e. $f_{i}(Q,T)$, $i=1,2,3$ for some chosen parameter space and show that the plots remain at the positive level. These are shown in Figs.(1), (2) and (3), where the functions are plotted against $Q$ and $T$ in 3D plots, for some chosen parameter space. We see that the functions remain at the positive level. This shows that for some suitable parameter space, the energy conditions are satisfied for this model. This exhibits the physical viability of the solutions obtained.

\subsubsection{Model-2}
Here, we consider the non-minimally coupled form
\begin{equation}
f(Q,T)=\lambda Q^{n}T^{m_{1}}
\end{equation}
where $\lambda$ is a constant. For this model, the inequalities in (\ref{td}) are reduced to
\begin{equation}
f_{4}(Q,T)=Q^{n}T^{m_{1}}\lambda\left(-1+n\left(1+\frac{2\left(-1+\frac{2T}{T+\lambda m_{1} Q^{n}T^{m_{1}}}\right)\omega^{2}}{Q}\right)\right)\geq 0
\end{equation}
\begin{figure}[hbt!]
\begin{center}
\includegraphics[height=3in]{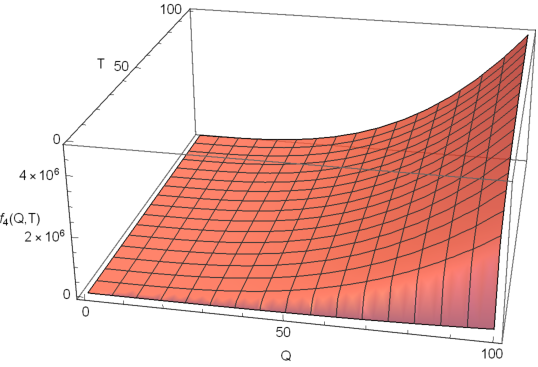}
\caption{The figure shows the plot of $f_{5}(Q,T)$ against $Q$ and $T$ for Model-2. Other parameters are given by $n=3, m_{1}=0.2, \lambda=1$.}
\label{figscale1}
\end{center}
\end{figure}
\begin{equation}
f_{5}(Q,T)=\frac{2\omega^{2}\lambda n Q^{n-1}T^{m_{1}}}{1+\lambda m_{1} Q^{n} T^{m_{1}-1} } \geq 0
\end{equation}

\begin{figure}[hbt!]
\begin{center}
\includegraphics[height=3in]{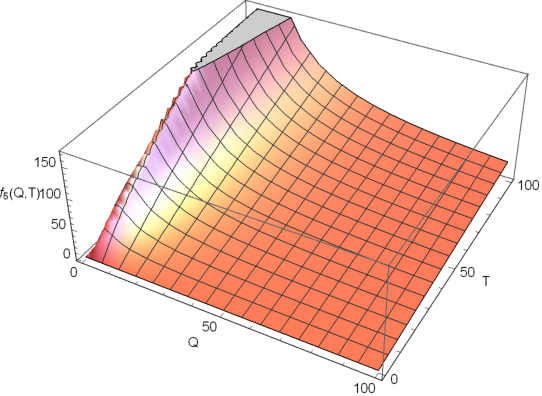}
\caption{The figure shows the plot of $f_{5}(Q,T)$ against $Q$ and $T$ for Model-2. Other parameters are given by $n=3, m_{1}=0.2, \lambda=1$.}
\label{figscale1}
\end{center}
\end{figure}

\begin{equation}
f_{6}(Q,T)=Q^{n} T^{m_{1}} \lambda \left(1+n\left(-1+\frac{2}{Q}\left(1+\frac{T}{T+\lambda m_{1}Q^{n}T^{m_{1}}}\right)\omega^{2}\right)\right)\geq 0
\end{equation}
\begin{figure}[hbt!]
\begin{center}
\includegraphics[height=3in]{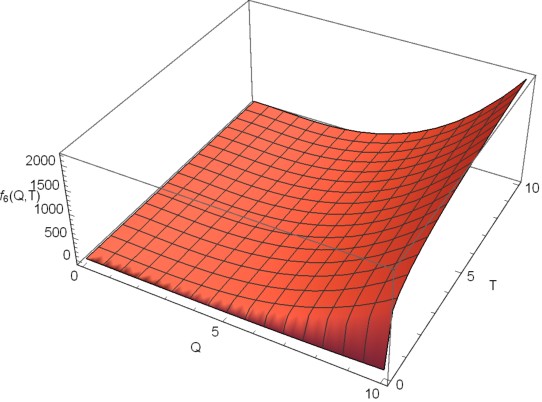}
\caption{The figure shows the plot of $f_{6}(Q,T)$ against $Q$ and $T$ for Model-2. Other parameters are given by $n=3, m_{1}=0.2, \lambda=1$.}
\label{figscale1}
\end{center}
\end{figure}

Just like the previous model, we generate plots for the functions $f_{i}(Q,T)$, $i=4,5,6$, to illustrate the validity of the energy conditions. These are shown in Figs.(4), (5) and (6), where the functions are plotted against $Q$ and $T$ in 3D plots, for some chosen parameter space. We see that the functions remain at the positive level. This shows that for some suitable parameter space, the energy conditions are satisfied for this model. This exhibits the physical viability of the solutions obtained.

\section{Causal Solutions with massless scalar field}
The most basic field that can exist in nature is a massless scalar field, which has no internal degrees of freedom, no spin, and just gives each point in spacetime a single numerical value. Despite its seeming simplicity, it is a fundamental model in general relativity (GR) and quantum field theory (QFT), offering a clean lab for investigating field–geometry interactions, quantization, and wave propagation. Causal solutions may result from a source other than a perfect fluid. Here we will add a massless scalar field $\phi$ to the perfect fluid. The energy-momentum tensor is given by
\begin{equation}\label{td1}
T_{AB}=(\rho+p)u_{A}u_{B}-p\eta_{AB}+\partial_{A}\phi\partial_{B}\phi-\frac{1}{2}\eta_{AB}\eta^{CD}\partial_{C}\phi\partial_{D}\phi
\end{equation}

\subsection{Scalar field depending on $x_{1}$ alone}
For simplicity, we consider a scalar field depending on only $x_{1}$. 
The non-zero components of the EMT are
\begin{equation}
T_{00}=\rho+\frac{\phi'(x_{1})^{2}}{2},  T_{11}=p+\frac{\phi'(x_{1})^{2}}{2},
T_{22}=T_{33}=p-\frac{\phi'(x_{1})^{2}}{2};
\end{equation}
For this set-up, the field equations (\ref{t4}) are written as
\begin{equation}\label{tt5}
\left(3\omega^{2}-m^{2}\right)f_{Q}+\frac{1}{2}(Qf_{Q}-f)=(1+f_{T})\rho+pf_{T}+\phi'(x_{1})^{2}
\end{equation}
\begin{equation}\label{tt6}
\omega^{2}f_{Q}-\frac{1}{2}(Qf_{Q}-f)=p+\phi'(x_{1})^{2}
\end{equation}
\begin{equation}
   \omega^{2}f_{Q}-\frac{1}{2}(Qf_{Q}-f)=p-\phi'(x_{1})^{2}
 \end{equation}
\begin{equation}\label{tt7}
\left(m^{2}-\omega^{2}\right)f_{Q}-\frac{1}{2}(Qf_{Q}-f)=p-\phi'(x_{1})^{2}
\end{equation}
From the above equations, we see that when $\phi'(x_{1})=0$, we get back the perfect fluid scenario discussed in section 5.

\subsection{A combination of cosmological constant and scalar field depending on $x_{3}$ alone}
Let us consider 
\begin{equation}
f(Q,T)=h(Q,T)-2\Lambda
\end{equation}
It is similar to modifying the field equations by adding a cosmological constant $\Lambda$ and assuming that $\phi$ relies exclusively on $x_{3}$ and that there is no fluid. If we put $\rho = p = 0$, the total energy momentum tensor (\ref{td1}) becomes the one for the massless scalar field, which is diagonal and has parts that are not connected to each other.
\begin{equation}
T_{00}=-T_{11}=-T_{22}=T_{33}=\phi'(x_{3})^{2}/2.
\end{equation}

The field equations (\ref{t4}) are given by
\begin{equation}\label{bh1}
\left(3\omega^{2}-m^{2}\right)h_{Q}+\frac{1}{2}\left(Qh_{Q}-h\right)+\Lambda=\frac{\phi'(x_{3})^{2}}{2}
\end{equation}
\begin{equation}\label{bh2}
\omega^{2}h_{Q}-\frac{1}{2}(Qh_{Q}-h)-\Lambda= -
\frac{\phi'(x_{3})^{2}}{2}
\end{equation}
\begin{equation}\label{bh3}
\left(m^{2}-\omega^{2}\right)h_{Q}-\frac{1}{2}(Qh_{Q}-h)-\Lambda=\frac{\phi'(x_{3})^{2}}{2} 
\end{equation}
Adding (\ref{bh1}) and (\ref{bh2}), we get
\begin{equation}
(4\omega^{2}-m^{2})h_{Q}=0
\end{equation}
which gives
\begin{equation}
m^{2}=4\omega^{2}
\end{equation}
Equation (\ref{bm}) state that if $m^{2}=4\omega^{2}$, then $x_{1c}=\infty$, in other terms,the solution is causal, and there is no critical radius. Using $ m^{2}=4\omega^{2}$, the equations (\ref{bh2}) and (\ref{bh3}) reduce to
\begin{equation}\label{bh5}
\omega^{2}h_{Q}-\frac{1}{2}(Qh_{Q}-h)-\Lambda= -
\phi'(x_{3})^{2}
\end{equation}
\begin{equation}\label{bh6}
3\omega^{2}h_{Q}-\frac{1}{2}(Qh_{Q}-h)-\Lambda=\phi'(x_{3})^{2} 
\end{equation}
\begin{equation}\label{ll1}
\Lambda=\frac{1}{2}(Qh_{Q}+h)
\end{equation}
Adding (\ref{bh1}) and (\ref{bh2}), we get
\begin{equation}\label{ll2}
\phi'(x_{3})^{2}=\frac{1}{2}Qh_{Q}
\end{equation}
From the above equations (\ref{ll1}) and (\ref{ll2}), we see that a suitably chosen function $h$ that satisfies both equations will produce a scenario for a viable Godel universe in the given set-up. Several trial solutions may be considered to check the validity of the above solutions, and it is expected that there will be different classes of functions that will produce these solutions. For example, if we consider $\phi'(x_{3})=\sqrt{\frac{2\Lambda-h}{2}}$, the above equations are satisfied. So, with such a choice existence of CTCs is a possibility in $f(Q,T)$ gravity with cosmological constant and scalar constant.

\section{Discussion and Conclusion}
In this work, the classical Godel solution from general relativity is extended into the framework of modified gravity theories based on non-metricity $Q$ and the trace of the energy-momentum tensor $T$ in the context of $f(Q,T)$ gravity. Moreover, the cosmological constant has been included in the setup in various places in the matter sector. Initially, we discuss the intricacies of $f(Q,T)$ gravity and the compatibility of the Godel solutions in the framework. Solutions are found for some specific models of $f(Q,T)$, thus showing the possibility of closed time-like curves in $f(Q,T)$. This is a direct extension of the idea of closed time-like curves in general relativity to non-metricity-based modified gravity. Einstein’s equations describe how matter and energy curve spacetime, but they do not demand that spacetime must always be globally causal. From this work, it seems to be true for $f(Q,T)$ gravity as well. We considered matter in the form of a perfect fluid and explored solutions that present a possible violation of causality. We also explored solutions with matter in the form of a massless scalar field. In all the cases, we got feasible solutions in the Godel metric. Energy conditions were investigated for the perfect fluid solutions to rule out the unphysical solutions. It was found that there exists a parameter space in which the energy conditions are satisfied. So, for chosen initial conditions, we will get scenarios where the solutions obtained for Godel spacetime in $f(Q,T)$ gravity are perfectly physically viable. This shows that the existence of closed timelike curves is perfectly feasible for $f(Q,T)$ gravity when suitable initial conditions are considered. 

In this work, we have explored the effects that the presence of non-metricity imposes on the possibility of physically viable Godel spacetime solutions in the background of $f(Q,T)$ gravity. The emergence of Godel-type metrics suggests that $f(Q,T)$ gravity preserves the potential of causality violation at the geometric level by admitting spacetimes with closed timelike curves, similar to General Relativity and some of its variations. The degree of causality violation appears to be theory-dependent, as the non-metricity–matter coupling directly affects the critical radius for CTC formation and the conditions for causal violation, in contrast to GR. Additionally, we discovered that the effective energy density and pressure are altered by the trace coupling term $T$, opening up new parameter regimes where Godel-type solutions either permit or prohibit CTCs. This suggests that compared to both $f(Q)$ and $f(R,T)$ theories, $f(Q,T)$ gravity offers a more comprehensive causal framework. In conclusion, our findings show that causal and acausal cosmic models can coexist depending on the form of $f(Q,T)$ and extend the class of Godel-type solutions into the domain of symmetric teleparallel gravity connected to matter. Future research could examine observational constraints, dynamical stability, and perturbative evolution on such rotating solutions, which could provide fresh perspectives on how spacetime geometry, non-metricity, and causality interact in modified gravity theories.

\section*{Acknowledgments}
PR acknowledges the Inter-University Centre for Astronomy and Astrophysics (IUCAA), Pune, India, for granting a visiting associateship. The authors acknowledge IUCAA for the local hospitality during a visit, when a portion of this work was done.

\section*{Data Availability Statement}

No data was generated or analyzed in this study.

\section*{Conflict of Interest}

There are no conflicts of interest.

\section*{Funding Statement}

There is no funding to report for this article.



\begin{thebibliography}{99}

\bibitem{GA} G.’t Hooft, M. Veltman, Ann. Inst. Henri Poincare, vol. XX, 69 (1974)

\bibitem{ghor} M. H. Goroff and A. Sagnotti, Phys. Lett. B 160, 81 (1985).

\bibitem{GA2} M. Veltman, ”Quantum theory of gravitation”, in: Les Houches, Session XXVIII, North-Holland, 1976 – Methods in Field Theory (eds. R. Balian, J. Zinn-Justin), p. 265-327.

\bibitem{Ae} A. Riess et al, Astron. J. 116, 1009 (1998), astro-ph/9805201.

\bibitem{KP} K. Stelle, Phys. Rev. D 16, 953 (1977).

\bibitem{VP} V. A. Kostelecky, Phys. Rev. D 69, 105009 (2004), hep-th/0312310

\bibitem{mg1} S. Nojiri, S. D. Odintsov, V. K. Oikonomou, Phys. Rept. 692 (2017) 1-104

\bibitem{mg2} C. G. Bohmer, E. Jensko, Phys. Rev. D., 104, 024010 (2021)

\bibitem{nab} Hans A. Buchdahl, Mon. Not. Roy. Astron. Soc., 150, 1 (1970).

\bibitem{nab1} J. Sadeghi, M. Shokri, S. N. Gashti, B. Pourhassan, P. Rudra, Int. J. Mod. Phys. D 31, 03, 2250019 (2022)

\bibitem{fg} T. Harko, Francisco S.N. Lobo, Shin'ichi Nojiri, Sergei D. Odintsov, Phys.Rev.D, 84, 024020 (2011)

\bibitem{fg1} P. Rudra, K. Giri, Nucl. Phys. B, 967, 115428 (2021)

\bibitem{fg2} P. Rudra, Int. J. Mod. Phys. D 31, 13, 2250095 (2022)

\bibitem{ct} A. Paliathanasis, J. D. Barrow, P. G. L. Leach, Phys. Rev. D, 94, 023525 (2016)

\bibitem{ct1} T. Ghorui, P. Rudra, F. Rahaman, Int. J. Mod. Phys. D, 34, 04, 2550014 (2025)

\bibitem{fq} J. B. Jimenez, L. Heisenberg, T. Koivisto, Phys. Rev. D., 98, 044048 (2018)

\bibitem{fqq} P. Saha, P. Rudra, Int. J. Mod. Phys. D 34, 03, 2550006 (2025)

\bibitem{fqttt} Y. Xu, G. Li, T. Harko, S-D. Liang, Eur. Phys. J. C., 79, 708 (2019)

\bibitem{fqttt1} A. Nájera, A. Fajardo, JCAP 03, 03, 020 (2022)

\bibitem{fqttt2} A. Nájera, A. Fajardo, Phys.Dark Univ. 34, 100889 (2021) 

\bibitem{KG} K. Godel, Rev. Mod. Phys. 21, 447 (1949).

\bibitem{sw}  S. Hawking, Phys. Rev. D 46, 603 (1992).

\bibitem{MJP}  M. Reboucas, J. Tiomno, Phys. Rev. D 28, 6 (1983); 

\bibitem{MJP1} M. Reboucas, M. Aman, A. F. F. Teixeira, J. Math. Phys. 27, 1370 (1985); 

\bibitem{MJP2} M. O. Galvao, M. Reboucas, A. F. F. Teixeira, W. M. Silva, Jr, J. Math. Phys. 29, 1127 (1988).

\bibitem{MDJ} M. Dabrowski, J. Garecki, Class. Quant. Grav. 19, 1 (2002)

\bibitem{jbm} J. Barrow, M. Dabrowski, Phys. Rev. D 58, 103502 (1998) 

\bibitem{jbm1} P. Kanti, C. E. Vayonakis, Phys. Rev. D 60, 103519 (1999)

\bibitem{ofn}  O. Bertolami, F. Lobo, NeuroQuantol. 7, 1 (2009), arXiv: 0902.0559 [gr-qc].

\bibitem{XSP} X. He, B. Wang, S. Chen, Phys. Rev. D 79, 084005 (2009),

\bibitem{nn1} A. M. Silva, M. J. Rebouças, N. A. Lemos, Int. J. Mod. Phys. D, 33, 16 (2450060) (2024)

\bibitem{nn2} A. J. C. Canuto, A. F. Santos, Eur. Phys. J. C 83, 5, 404 (2023)

\bibitem{nn3} B. K. Bishi, P. V. Lepse, A. Beesham, Chin. J. Phys., 81, 162 (2023)

\bibitem{nn4} J. S. Gonçalves, A. F. Santos, Eur. Phys. J. C, 82, 11, 979 (2022).

\bibitem{RIE} R. D’Inverno, Introducing Einstein’s Relativity, (Oxford University Press, 1992) 145-152.

\bibitem{godd1} M. J. Rebouças, J. E. Aman, J. Math. Phys. 28, 888 (1987)

\bibitem{l1} J. Santos, M. J. Reboucas, A. F. F. Teixeira, Eur. Phys. J. C 78, 567 (2018)

\bibitem{MJR} M. J. Rebou¸cas, Modelos do Universo com Rota¸c˜ao Dependente do Tempo e Viola¸c˜ao da Causalidade na Cosmologia, PhD Thesis (in Portuguese). Centro Brasileiro de Pesquisas Fısicas, Rio de Janeiro, Brasil (1981)

\bibitem{JMS} J. Santos, M. J. Rebou¸cas and A. F. F. Teixeira, Eur. Phys. J. C 78, 567 (2018).

\bibitem{en1} S. W. Hawking and G. F. R. Ellis, The Large Scale Structure of Spacetime (Cambridge University Press, Cambridge, 1973).

\bibitem{en2} S. Carroll, Spacetime and Geometry (Addison-Wesley, New York, 2004).


\end{thebibliography}
\end{document}